\documentclass[aps,twocolumn,floats,prd,nofootinbib,10pt,floatfix]{revtex4-1}

\usepackage[pdftitle=tSZ-ISW,pdfauthor=Simeon Bird]{hyperref}
\usepackage{graphicx,amsmath,amsfonts,amssymb,slashed}
\usepackage{bbold,wasysym}

\usepackage[usenames,dvipsnames]{xcolor} 

\usepackage{soul}

\definecolor{RedWine}{rgb}{0.743,0,0}
\definecolor{RoyalBlue}{rgb}{0.25,.41,.88}

\setstcolor{Blue}

\def\VEV#1{\left\langle #1 \right\rangle}

\begin{document}

\title{Cross-correlation between thermal Sunyaev-Zeldovich
     effect and the integrated Sachs-Wolfe effect}
\author{Cyril Creque-Sarbinowski$^1$, Simeon Bird$^2$, and Marc
     Kamionkowski$^2$} 
\affiliation{$1$Department of Physics, Massachusetts Institute
     of Technology, 77 Massachusetts Avenue, Cambridge, MA 02139, USA\\
     $^2$Department of Physics and Astronomy, Johns Hopkins
     University, 3400 N.\ Charles Street, Baltimore, MD 21218, USA}

\begin{abstract}
Large-angle fluctuations in the cosmic microwave background
(CMB) temperature induced by the integrated Sachs-Wolfe (ISW)
effect and Compton-$y$ distortions from the thermal
Sunyaev-Zeldovich (tSZ) effect are both due to line-of-sight
density perturbations.  Here we calculate the cross-correlation
between these two signals.  Measurement of this
cross-correlation can be used to test the redshift distribution
of the tSZ distortion, which has implications for the redshift at which 
astrophysical processes in clusters begin to operate.  We also evaluate 
the detectability of a $yT$ cross-correlation from exotic early-Universe 
sources in the presence of this late-time effect.
\end{abstract}

\maketitle
\section{Introduction}
\label{sec:intro}

The standard $\Lambda$CDM cosmological model provides a
remarkably good fit to an array of precise measurements.
However, there still remain some tensions between different
measurements which must be resolved, and the physics responsible
for the generation of primordial perturbations has yet to be
delineated.  This paper addresses both these issues.

Large-angle fluctuations in the cosmic microwave background
(CMB) temperature ($T$) are induced not only by density perturbations
at the CMB surface of last scatter (the Sachs-Wolfe effect; SW), but
also by the growth of density perturbations along the line of
sight (the integrated Sachs-Wolfe effect; ISW)
\cite{Sachs:1967er}.  Although the
CMB frequency spectrum is very close to a blackbody, there are
small distortions, of the Compton-$y$ type, induced by the rare
scattering of CMB photons from hot
electrons in the intergalactic medium (IGM) of galaxy clusters
\cite{Sunyaev:1972eq}.  This $y$ distortion has been mapped, as
a function of position on the sky, by Planck with an angular
resolution of a fraction of a degree
\cite{Ade:2013qta,Aghanim:2015eva}, and there are vigorous
discussions of future missions, such as PIXIE \cite{Kogut:2011xw}
and PRISM \cite{Andre:2013nfa}, that will map the $y$ distortion
with far greater sensitivity and resolution.

Given that both the tSZ and ISW fluctuations are induced by
density perturbations at relatively low redshifts, there should
be some cross-correlation between the two \cite{Taburet:2010hb},
and the purpose of
this paper is to calculate this cross-correlation.  The
motivation for this work is two-fold:  First, there is some
tension between the measured amplitude of $y$ fluctuations and
the amplitude of density perturbations inferred from CMB
measurements
\cite{Lueker:2009rx,Komatsu:2010fb,Ade:2013lmv,Ade:2015fva}.
The tension, though, is based
upon theoretical models that connect the $y$-distortion and
density-perturbation amplitudes.  Ingredients of these models
include nonlinear evolution of primordial perturbations, gas
dynamics, and feedback processes, all of which can become quite
complicated.  Any empirical handle on this physics would
therefore be useful. To quantify how well the cross-correlation
can constrain these processes, we introduce a parameter, 
$\epsilon$, which describes the peak redshift of the cross-correlation 
signal. If clusters were to develop a hot envelope earlier than currently 
expected from theory, $\epsilon$ would increase. We design this 
parameter so that it does not affect the tSZ signal, merely the 
cross-correlation. Using our formalism we quantify
how well the cross-correlation breaks the degeneracy between 
structure formation parameters (such as the amplitude of 
fluctuations, $\sigma_8$) and the astrophysical processes which
lead to the halo pressure profile.

The second motivation involves the search for exotic
early-Universe physics.  Recent work has shown that primordial
non-gaussianity may lead to a $yT$ cross-correlation
which may be used to probe scale-dependent non-gaussianity
\cite{Emami:2015xqa}.  The present calculation will be used to
explore whether this early-Universe signal can be distinguished
from late-time effects that induce a $yT$ correlation.

This paper is organized as follows.  In Section
\ref{sec:calculation} we derive expressions for the power
spectra for the ISW effect, the tSZ effect, and their
cross-correlation, and then present numerical results.  In
Section \ref{sec:sz} we evaluate the prospects to infer some
information about the redshift distribution for tSZ fluctuations
from measurement of the tSZ-ISW cross-correlation.  In
Section \ref{sec:ng} we estimate the sensitivity of future
measurements to the tSZ-ISW cross-correlation from primordial
non-gaussianity.  We conclude in Section \ref{sec:concl}.

\section{Calculation}
\label{sec:calculation}

\subsection{The ISW Effect}

The integrated Sachs-Wolfe (ISW) effect describes the frequency
shift of CMB photons as they traverse through time-evolving
gravitational potentials.  The fractional temperature
fluctuation in a direction $\hat n$ due to this frequency shift
is 
\begin{equation}
     \frac{\Delta T}{T}(\hat{n}) =
     -\frac{2}{c^2}\int d\eta\,
     \frac{d\phi}{d\eta}(c\eta \hat n;z)
     = -\frac{2}{c^2}\int\, dz 
     \frac{d\phi}{dz}(r \hat n;z),
\label{eqn:isw}
\end{equation}
where $\phi(\vec x,z)$ is the gravitational potential at
position $\vec x$ and redshift $z$, $\eta$ the conformal time,
$c$ the speed of light, and $r$ the distance along the line of
sight.

The potential $\phi$ is related to the density perturbation
through the Poisson equation $\nabla^2 \phi = 4 \pi G \rho$,
where $\nabla$ is a gradient with respect to physical position,
$G$ Newton's constant, and $\rho$ the matter density. We write
$\rho(\vec x;z)= \bar\rho [1+\delta(\vec x;z)]$ in terms of the
mean matter density $\bar \rho$ and fractional density
perturbation $\delta(\vec x;z)$.  We then use the Friedmann
equation to write $\bar \rho = (3/8\pi G) \Omega_m H_0^2 a^{-3}$ in
terms of the matter density $\Omega_m$ (in units of the critical
density), Hubble parameter $H_0$, and scale factor
$a=(1+z)^{-1}$.  We further write the density perturbation
$\delta(\vec x;z) = D(z) \delta(\vec x;z=0)$ in terms of the
linear-theory growth factor $D(z)$.  We can then re-write the
Poisson equation as
\begin{equation}
     \phi(\vec x;z) = -\frac{3}{2} \Omega_m H_0^2
     \frac{D(z)}{a(z)} \left[ \nabla_c^{-2} \delta(\vec x;z=0)
     \right],
\label{eqn:phik}
\end{equation}
where $\nabla_c=\nabla/a$ is the gradient with respect to
the comoving coordinates.

The power spectrum for ISW-induced angular temperature
fluctuations is then obtained using the Limber approximation,
which can be stated as follows:  If we observe a two-dimensional
projection,
\begin{equation}
     p(\hat n) = \int \, dr\, q(r) \delta(r \hat n),
\end{equation}
of a three-dimensional field $\delta(\vec x)$, with
line-of-sight-distance weight function $q(r)$, then the angular
power spectrum, for multipole $l$, of $p(\hat n)$ is
\begin{equation}
    C_l^p = \int \, dr\, \frac{[q(r)]^2}{r^2} P\left(\frac{l+1/2}{r}\right),
\label{eqn:Limber}
\end{equation}
in terms of the three-dimensional power spectrum $P(k)$, for
wavenumber $k$, for $\delta(\vec x)$.

Using Eqs.~(\ref{eqn:isw}), (\ref{eqn:phik}), and
(\ref{eqn:Limber}), we find the power spectrum for ISW-induced
temperature fluctuations to be,
\begin{eqnarray}
     C_l & = & \left(\frac{3 \Omega_m H_0^2}{c^3 (l+1/2)^2} \right)^2
     c \int r^2  dr \left[ H(z) \frac{d}{dz}
     \left( \frac{D(z)}{a(z)} \right) \right]^2 P\left(k
     \right) \nonumber \\ 
        & = & \left(\frac{3 \Omega_m H_0^2}{c^3(l+1/2)^2} \right)^2
     \int  c\,dz H(z) \left[r(z) \frac{d}{dz}
     \left( \frac{D(z)}{a(z)} \right) \right]^2 P\left(k
     \right) \nonumber \\
     & = &  \int \frac{c\, dz}{H(z)} \left[ \Delta^{\rm
     isw}_l(z)\right]^2 P\left(\frac{l+1/2}{r}\right),
\label{eqn:Clisw}
\end{eqnarray}
in terms of the matter-density power spectrum $P(k)$ today.
Note that we used the relation $dz = -(H/c) dr$ to get from the
first to the second line in Eq.~(\ref{eqn:Clisw}), and we have
defined in the last line the ISW transfer function,
\begin{equation}
     \Delta^{\rm isw}_l(z) = \frac{3 \Omega_m H_0^2}{c^3 (l+1/2)^2}
     r(z) H(z) \frac{d}{dz}
     \left( \frac{D(z)}{a(z)} \right).
\label{eqn:iswtransfer}
\end{equation}

\subsection{The Thermal Sunyaev-Zeldovich Effect}

The thermal SZ effect (tSZ) arises from inverse-Compton
scattering from the hot electrons in the intergalactic medium of
galaxy clusters. This upscattering induces a frequency-dependent
shift in the CMB intensity in direction $\hat n$ which we write
as a brightness-temperature fluctuation,
\begin{align}
     \left(\frac{\Delta T}{T}\right)_\nu(\hat{n}) = g(\nu) y \equiv
     \left(x\frac{e^x + 1}{e^x - 1} - 4\right)y(\hat{n}),
\end{align}
where $y(\hat n)$ is the $y$ distortion in direction $\hat n$,
and $x \equiv h\nu/k_B T$, with $\nu$ the frequency, $k_B$ the
Boltzmann constant, $h$ the Planck constant, and $T=2.7255$~K the
CMB temperature \citep{Fixsen:2009}.  The Compton-$y$ distortion is given by an integral,
\begin{align}
     y(\hat{n}) &\equiv \frac{k_B\sigma_T}{m_e
     c^2}\int\,  ds \, n_e(s \hat n) T_e
     (s \hat n),
\label{eqn:yparameter}
\end{align}
along the line of sight, where $s$ is the (physical) line-of-sight
distance, $\sigma_T$ the Thomson cross section, $n_e(\vec x)$ the
electron number density at position $\vec x$, and $T_e(\vec x)$
the electron temperature.  The hot electrons that give rise to
this distortion are assumed to be housed in galaxy clusters
with a variety of masses $M$ and a variety of redshifts $z$.
The spatial abundance of clusters with masses between $M$ and
$M+dM$ at redshift $z$ is $(dn/dM)dM$ in terms of a mass
function $(dn/dM)(M,z)$, a function of mass and redshift.
Galaxy clusters of mass $M$ at redshift $z$ are distributed
spatially with a fractional number-density perturbation that is
assumed to be $b(M,z)\delta(\vec x)$ in terms of a bias
$b(M,z)$.  The spatial fluctuations to the electron pressure
$P_e(\vec x)=k_B n_e(\vec x) T_e(\vec x)$ that give rise to
angular fluctuations in the Compton-$y$ parameter induced by
clusters of mass $M$ and redshift $z$ can then be
modeled as $b(M,z)$ times a convolution of the density
perturbation $\delta(\vec x)$ with the electron-pressure profile
of the cluster.  Since convolution in configuration space
corresponds to multiplication in Fourier space, the Limber
derivation discussed above can be used to find the power
spectrum for angular fluctuations in the Compton-$y$ parameter
to be
\cite{Komatsu:1999ev,Diego:2004uw,Taburet:2010hb,Ade:2013qta,Ade:2015mva},
\begin{equation}
     C_l^{yy,2h} = \int \frac{c\, dz}{H(z)} \left[
     \Delta^{y}_l(z) \right]^2
     P\left(\frac{l+1/2}{r}\right),
\label{eqn:Cly2h}
\end{equation}
in terms of a transfer function,
\begin{equation}
     \Delta_l^{y}(z) = r(z) D(z) \int \frac{dn}{dM}dM 
     y_l(M,z) b(M,z).
\end{equation}
Here, $y_l(M,z)$ is the 2d Fourier transform of the
Compton-$y$ image, on the sky, of a cluster of mass $M$ at
redshift $z$ and is given in terms of the electron pressure
profile $P_e(M,z;x)$, as a function of scale radius $x$ in the
cluster. We neglect relativistic effects, which are second-order 
for our purposes \cite{Itoh:1997ks}.
We use for our numerical work the electron-pressure profiles of
Ref.~\cite{Nagai:2007mt, Arnaud:2009tt} with the parameters 
described in \cite{Dolag:2015dta}. We assume the halo mass 
function of Ref.~\cite{Tinker:2008ff} and the halo bias of
Ref.~\cite{Sheth:1999mn}.

The ``2h'' superscript in the
$y$-parameter power spectrum indicates that this is the
``two-halo'' contribution, the $y$ autocorrelation that arises
from large-scale density perturbations.  There is an additional
``one-halo'' contribution that arises from Poisson fluctuations
in the number of clusters.  This is
\cite{Komatsu:2002wc},
\begin{equation}
     C_l^{yy,1h} = \int\, dz \left[r(z)\right]^2\frac{c}{H(z)}
     \int dM \frac{dn(M,z)}{dM} \left| y_l(M,z)
     \right|^2.
\label{eqn:Cly1h}
\end{equation}
The total $y$-parameter power spectrum is $C_l^{yy} =
C_l^{yy,1h}+C_l^{yy,2h}$.  To avoid our signal being dominated by
unphysical $z\sim 0$ objects, we place a lower integration limit 
of $z=0.02$, the redshift of the COMA cluster.

\subsection{ISW-tSZ cross-correlation}

Given that the temperature fluctuation induced by the ISW effect
and the two-halo contribution to tSZ fluctuations are both
generated on large angular scales by the same fractional density
perturbation $\delta(\vec x)$, there should be a
cross-correlation between the two.  From the expressions,
Eq.~(\ref{eqn:Clisw}) and (\ref{eqn:Cly2h}), it is clear that this
cross-correlation is
\begin{equation}
     C_l^{yT} = \int \frac{c\, dz}{H(z)} \Delta^{\rm isw}_l(z)
     \Delta^y_l(z) P\left(\frac{l+1/2}{r}\right).
\label{eqn:Clcross}
\end{equation}

\begin{figure}[h]
\includegraphics[width=9.5cm, height = 7 cm]{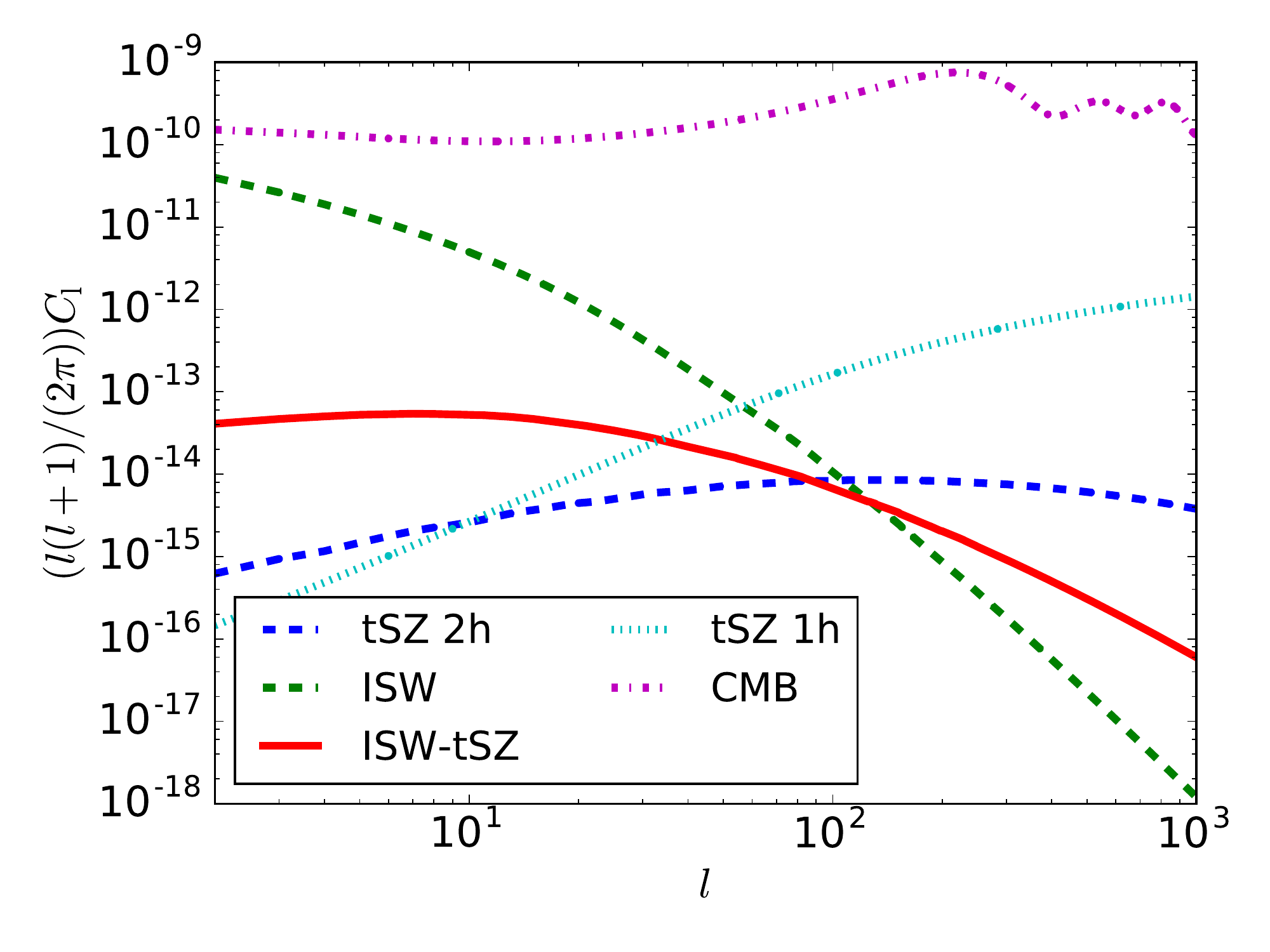}
    \caption{The ISW power spectrum $C_l^{\rm isw}$ (green) and
    the two-halo contribution ($y,2h$) to $C_l^{yy}$ (blue) are
    shown in dashed
    lines, while $C_l^{yT}$ is shown in solid
    red. The one-halo contribution ($y, 1h$) to $C_l^{yy}$ is
    dotted.  The CMB power spectrum is shown dot-dashed for
    comparison.} 
\label{fig:power}
\end{figure}

\subsection{Numerical results and approximations}

Fig.~\ref{fig:power} shows the resulting power spectra.  
For our numerical results, we use a vacuum-energy density (in
units of critical) $\Omega_{\Lambda} = 0.721$, matter density
$\Omega_m = 0.279$, baryon density $\Omega_b = 0.046$, a
critical density for collapse of $\delta_c = 1.686$, and
dimensionless Hubble parameter $h= 0.701$, although the
large-angle results that will be our primary focus are largely
insensitive to these details. In practice, we find that over ninety percent 
of the contribution comes from the redshift range $z = 0-3$ and the halo mass range 
$10^{13}\,M_\odot$ and $10^{15}\,M_\odot$. 
We therefore integrate over slightly wider ranges; $z = 0.02 - 4$  
and halo mass $10^{12}\,M_\odot$ and $10^{16}\,M_\odot$.

The large angle (low-$l$) behaviors of the ISW-ISW
autocorrelation, the $yT$ cross-correlation, and the one- and
two-halo contributions to the $yy$ power spectra are easy to
understand qualitatively.  Let us begin with the ISW effect.  Here, the
$l$ dependence of the transfer function is $\Delta^{\rm isw}_l
\propto l^{-2}$, and for large angles ($l\lesssim 20$), the power
spectrum is $P(l/r) \propto l$, assuming $l+1/2 \approx l$.  As a result, $l^2C_l^{\rm isw}
\propto l^{-1}$ for $l \lesssim 20$.  Next consider the tSZ
power spectra.  Galaxy clusters subtend a broad distribution of
angular sizes but are rarely wider than a degree.  Thus, for 
$l \lesssim 20$, they are effectively point sources.  The
Fourier transform is thus effectively approximated by $y_l(M,z)
\simeq y_{l=0}(M,z)$ which is itself precisely the integral of
the $y$-distortion over the cluster image on the sky, or
equivalently, the total contribution of the cluster to the
angle-averaged $y$.  As a result of the independence of $y_l$ on
$l$ and $P(l/r) \propto l$ for $l \lesssim 20$, we infer
$l^2 C_l^{yy,2h} \propto l$ and $l^2 C_l^{yT} \propto $~const for
$l\lesssim 20$.  Finally, the one-halo contribution to
$C_l^{yy}$ is nearly constant (i.e., $l^2C_l\propto l^2$) for
$l\lesssim 20$ as expected for Poisson fluctuations in what are
(at these angular scales) effectively point sources.

\section{SZ redshift distribution}
\label{sec:sz}

\begin{figure}[h]
\includegraphics[width=9.5cm, height = 7 cm]{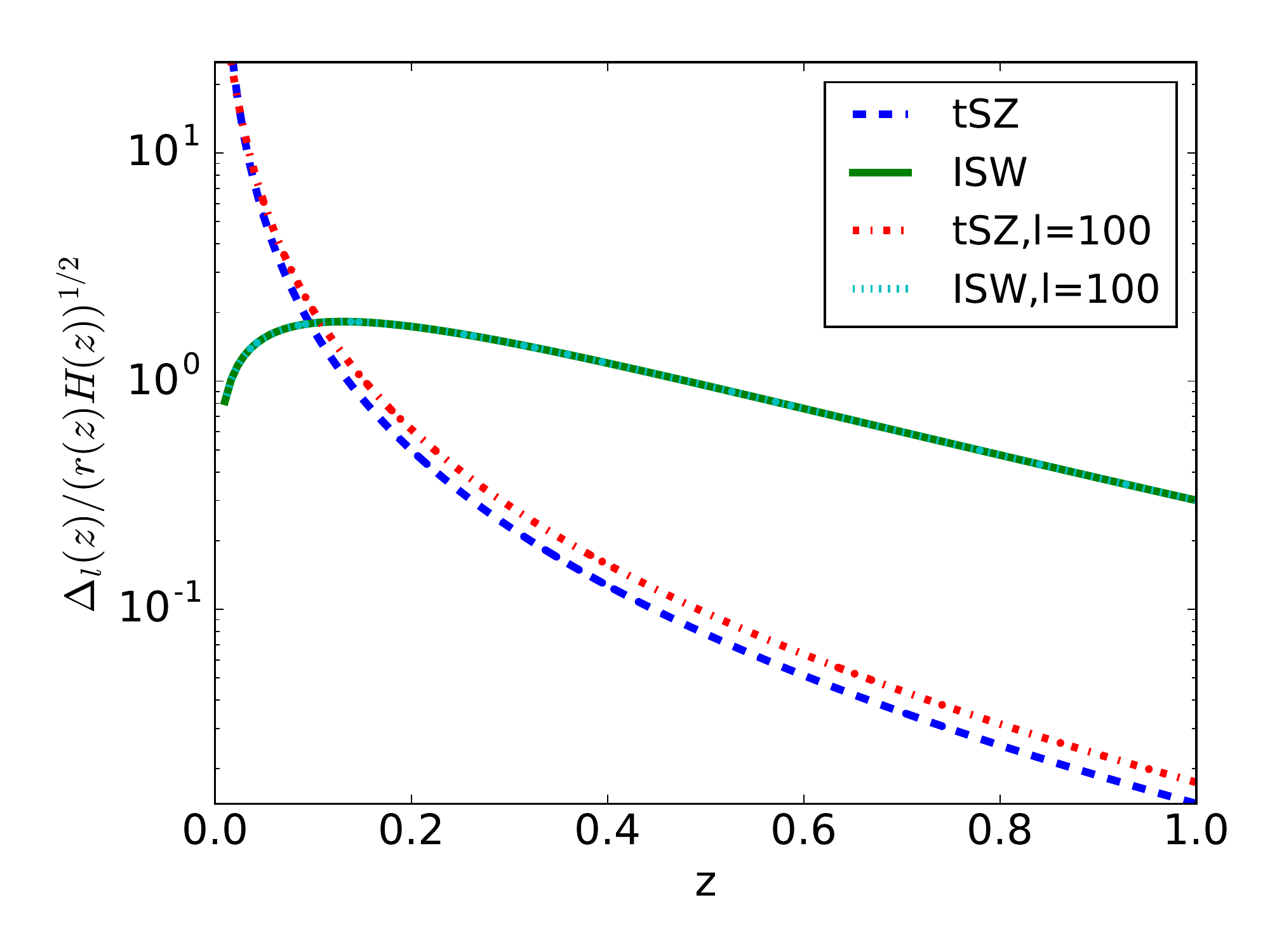}
\caption{We plot the transfer functions $\Delta_l^{\rm isw}(z)$
     and $\Delta_l^y(z)$, divided by $[r(z)H(z)]^{1/2}$, at $l=20$. The
     squares of the plotted quantities are the redshift ($z$)
     integrands for the ISW power spectrum $C_l^{\rm isw}$ and
     the two-halo contribution to the tSZ power spectrum
     $C_l^{yy}$.  Both curves are normalized so that the
     areas under the curve are the same.  The tSZ-ISW
     cross-correlation $C_l^{yT}$ is obtained from the overlap
     of these two.}
\label{fig:Deltas}
\end{figure}

We now discuss the prospects to learn about the redshift
distribution of the galaxy clusters that produce the Compton-$y$
distortion.  As seen above, the $yT$ correlation is significant
primarily at multipole moments $l \lesssim 100$, where the
window function $y_l(z)$ is largely independent of $l$.  The
amplitude of the cross-correlation, relative to the
auto-correlations, can be largely understood by examining the
overlap between the redshift dependences of the two transfer
functions $\Delta_l^y(z)$ and $\Delta_l^{\rm isw}(z)$.  These
transfer functions are shown in Fig.~\ref{fig:Deltas}.  More
precisely, we plot--noting that $P(l/r) \propto l/r$ for the
relevant angular scales---$\Delta_l/\left[H(z) r(z) \right]^{1/2}$,
the square root of the integrands for $C_l$, as it is the
overlap of these two functions that determines the strength of
the cross-correlation relative to the auto-correlation.  We also
normalize the curves in Fig.~\ref{fig:Deltas} to both have the
same area under the curve.

Given the current fairly precise constraints to dark-energy
parameters, the predictions for $\Delta_l^{\rm isw}(z)$ has
relatively small uncertainties.  The prediction for
$\Delta_l^y(z)$ depends, however, on the redshift distribution
of the halo mass function, bias parameters, and cluster pressure
profiles, all of which involve quite uncertain physics.
Measurement of the $yT$ correlation will, however, provide an
additional empirical constraint on the redshift evolution of the
$y$ parameter.

To see how this might work, we replace
\begin{equation}
     \Delta_l^y(z) \to \Delta_l^y(z) \left[ 1 +
     \epsilon (z-z_0) \right],
\label{eqn:replacement}
\end{equation}
where
\begin{equation}
     z_0 = \frac{ \int \frac{dz}{r(z)H(z)} z \Delta_l^y(z)}
     { \int \frac{dz}{r(z)H(z)} \Delta_l^y(z)} \simeq 0.04\left(\frac{l}{100}\right).
\label{eqn:z0}     
\end{equation}
The functional form in Eq.~(\ref{eqn:replacement}), is chosen so
that, with $z_0$ given in Eq.~(\ref{eqn:z0}), the
auto-correlation power spectrum $C_l^{yy}$ will remain unaltered
for small $\epsilon$.  This alteration thus describes, for
$\epsilon>0$, a weighting of the Compton-$y$ distribution to
smaller redshifts (and {\it vice versa} for $\epsilon<0$) in
such a way that leaves the total $y$ signal unchanged.

\begin{figure}[h]
\includegraphics[width=9.5cm, height = 7 cm]{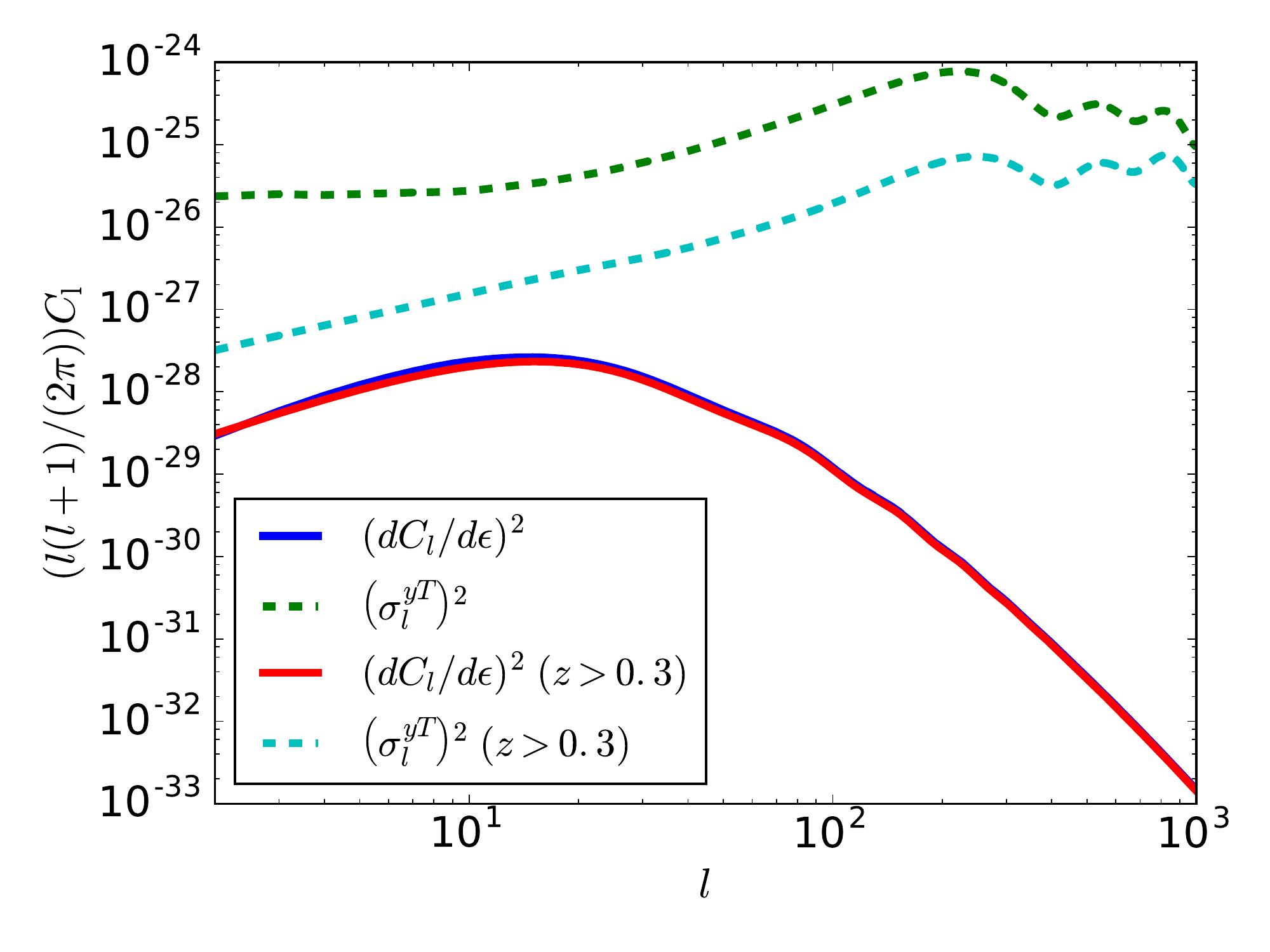}
\caption{The derivative $\partial C_l^{yT}/\partial \epsilon$ of
    the ISW-tSZ cross-correlation with respect to $\epsilon$
    (solid blue), and the ISW-tSZ noise $\sigma_l^{yT}$ (dashed
    green).  We also show the same quantities restricting to
    $z>0.3$ (solid red and dashed cyan, respectively),  
    which substantially increases the signal-to-noise.} 
\label{fig:clderiv}
\end{figure}

We now estimate the smallest value $\sigma_\epsilon$ of
$\epsilon$ that will be detectable with future measurements.
This is given by
\begin{equation}
     \frac{1}{\sigma_\epsilon^2} \simeq \sum_l \frac{ \left( \partial
     C_l^{yT}/\partial \epsilon \right)^2}{\left(\sigma_l^{yT} \right)^2},
\end{equation}
where
\begin{equation}
     \frac{\partial C_l^{yT}}{\partial \epsilon} = \int
     \frac{c\, dz}{H(z)} \Delta^{\rm isw}_l(z) \Delta^y_l(z)
     (z-z_0) P(l/r).
\label{eqn:Clderiv}
\end{equation}
Fig.~\ref{fig:clderiv} shows $\partial C_l^{yT}/\partial
\epsilon$ and $\sigma_l^{yT}$. 

The error with which each $C_l^{yT}$ can be determined is
\begin{equation}
     \left(\sigma_l^{yT} \right)^2 =\frac{1}{2l+1} \left[ \left(
     C_l^{yT} \right)^2 + C_l^{TT}\left(
     C_l^{yy} + N_lW_l^{-2} \right)\right],
\label{eqn:noise}
\end{equation}
where $C_l^{TT}$ is the CMB temperature power spectrum,
$W_l = e^{-l^2\sigma_b^2/2}$ is a window function, and $ N_l
= (4\pi/N)\sigma_y^2$ is the noise in the measurement of
$C_l^{yy}$.  Here, $\sigma_b$ the beam size and, $\sigma_y$ the
root-variance of the $y$-distortion measurement in each
pixel, and $N$ the number of pixels.

The Planck satellite has now measured the tSZ power spectrum and
found good agreement with the expectations from the one-halo
contribution to $C_l^{yy}$.  They have now even presented good
evidence for detection of the two-halo contribution at
$l\lesssim 10$.  From this we infer that the noise contribution
$N_l$ to the $yy$ measurement is already small compared with
$C_l^{yy}$, and it will be negligible for future experiments
like PIXIE or PRISM.  We also note from the numerical results
that $\left(C_l^{yT}\right)^2$ is small compared with $C_l^{yy}
C_l^{TT}$---this makes sense given that the cross-correlation of
$y$ with the ISW effect is small and further that the ISW effect
provides only a small contribution to large-angle temperature
fluctuations.  We may thus approximate
\begin{equation}
     \left(\sigma_l^{yT} \right)^2 \simeq \frac{1}{2l+1}
     C_l^{TT} C_l^{yy}.
\label{eqn:noiseapprox}     
\end{equation}

The smallest detectable value of $\epsilon$ evaluates to
$\sigma_\epsilon \simeq 2.3$, for $z_0 = 0.13$. 
This is still a considerable uncertainty, but the signal to noise can be
increased.  While the majority of the cross-correlation signal
is at low redshifts, the 1-halo tSZ term, which appears as noise
in Eq.~(\ref{eqn:noise}), peaks even more strongly at low
redshift. Thus the lowest-redshift bins are substantially noise
dominated.\footnote{Eq.~\ref{eqn:noise} assumes Gaussian fluctuations for $C_l^{yy}$. 
This assumption is invalid at $z\sim 0$, where a single nearby large 
cluster can dominate large angular scales. The $z\sim 0$ signal is 
already extremely small, however, so this does not change our conclusions.}
If we remove all information at redshifts $z < 0.3$,
possible by explicitly detecting resolved clusters and masking
them from the tSZ map, then the noise can be considerably
reduced.  In this case, the smallest detectable value becomes
$\sigma_\epsilon \simeq 0.57$, which, if achieved, would
provide some valuable information on the redshift distribution
of tSZ fluctuations, constraining the formation of massive clusters 
to the era of dark energy dominance.

We also considered the profile of Ref.~\cite{Komatsu:2002wc}, 
which produces a larger tSZ signal and thus a larger cross-correlation. 
However, since the 1-halo tSZ term dominates the noise, this actually 
slightly decreased the sensitivity. Thus our results should be largely
insensitive to the electron pressure profile used.

\section{Primordial Non-Gaussianity}
\label{sec:ng}

We now review the $yT$ cross-correlation from the
scale-dependent primordial non-gaussianity scenario of
Ref.~\cite{Emami:2015xqa}.  If primordial perturbations are
non-gaussian, the amplitude of small-wavelength power can be
modulated by long-wavelength Fourier modes of the density
field.  The dissipation of primordial Fourier modes with
wavenumbers $k\simeq 1-50$ Mpc$^{-1}$ (which takes place at
redshifts $1100 \lesssim z \lesssim 5\times 10^4$) gives rise to
primordial Compton-$y$ distortions.  If there is non-gaussianity,
then the angular distribution of this $y$ distortion may be
correlated with the large-scale density modes that give rise,
through the Sachs-Wolfe effect, to large-angle fluctuations in
the CMB temperature.

The predictions for this primordial $yT$ correlation depend on the
yet-unmeasured isotropic value $\VEV{y}$ of the Compton-$y$
parameter for which we take as a canonical value $4\times
10^{-9}$.  The $yy$ and $yT$ power spectra for the scenario are
then, 
\begin{eqnarray}
     l^2 C_l^{yy,{\rm ng}} &\simeq& 5.5\times10^{-20} \, \left( \frac{
     f_{\rm nl}^y}{200} \right)^2 \left(\frac{\VEV{y}}{4\times
     10^{-9}} \right)^2, \\
     l^2 C_l^{yT,{\rm ng}} &\simeq& 5.8\times10^{-15} \, \left( \frac{
     f_{\rm nl}^y}{200} \right) \left(\frac{\VEV{y}}{4\times
     10^{-9}} \right).
\end{eqnarray}
Here, $f_{\rm nl}^y$ is the non-gaussianity parameter for
squeezed bispectrum configurations in which the wavenumber of
the long-wavelength mode is of the $\sim$Gpc$^{-1}$ scales of
modes that contribute to the ISW effect, while the two
short-wavelength modes have wavelengths $1\,{\rm Mpc}^{-1}
\lesssim k \lesssim 50\,{\rm Mpc}^{-1}$
As discussed in Ref.~\cite{Emami:2015xqa}, there are no
existing model-independent constraints on $f_{\rm nl}^y$.

We now estimate the detectability of the $yT$ cross-correlation
from non-gaussianity, discussed in Ref.~\cite{Emami:2015xqa}.
In that work, the late-time contribution to $C_l^{yy}$ and
$C_l^{yT}$ was neglected, and the detectability of the
primordial signal inferred assuming that detection of $y$
fluctuations was noise-limited.  Here we re-do those estimates
taking into account the late-time $yT$ correlation calculated
above.

If the late-time $yT$ is somehow known precisely, the
signal-to-noise with which an early-Universe $yT$ signal with
power spectrum $C_l^{yT,{\rm ng}}$ can be distinguished from the
null hypothesis is
\begin{equation}
     \left(\frac{S}{N} \right) = \left(\sum_l\frac{\left(C_l^{yT,{\rm ng}}
     \right)^2}{(\sigma_l^{yT})^2}\right)^{1/2}.
\label{eqn:SN}
\end{equation}
Using Eq.~(\ref{eqn:noiseapprox}) and the numerical results for
$C_l^{yy}$, we then obtain a signal-to-noise $(S/N) \simeq  (f_{\rm
nl}^y/1065) (\VEV{y}/4\times 10^{-9})$.  This calculation differs from 
that of Ref.~\cite{Emami:2015xqa} in two respects; we have included
the late-time contribution to Compton-$y$ fluctuations, which
degrades the detectability $f_{\rm nl}^y$ by about a factor of 4,
even if the late-time $yT$
correlation is assumed to be known precisely.  The
detectability is, moreover,
limited by cosmic variance and not from measurement noise.
We have included in the sum in Eq.~(\ref{eqn:SN}) angular
modes up to $l \leq 1000$; the signal-to-noise improves if the
sum is extended to higher $l$.

This calculation overestimates the smallest detectable
signal, as there is a theoretical uncertainty in the late-time
$yT$ correlation, as discussed in Section \ref{sec:sz}; it must
instead be determined from the data.  There is thus an
additional uncertainty to the inferred value of $f_{\rm nl}^y$
that will arise after marginalizing over the uncertain late-time
$yT$ amplitude.  We thus assume that the total $yT$ power
spectrum is a combination $C_l^{yT,{\rm tot}} = \alpha C_l^{yT}+
C_l^{yT,{\rm ng}}$ of the late-time and non-gaussian
contributions.  Here $\alpha \sim 1$ accounts for uncertainty in
the amplitude of $yT$.
We then calculate the Fisher matrix
\cite{Jungman:1995bz},
\begin{equation}
     F_{ij} = \sum_\ell \frac{\left( \partial C_l^{yT,{\rm
     tot}}/\partial s_i \right) \left( \partial C_l^{yT,{\rm tot}}/\partial
     s_j\right)}{\left(\sigma_l^{yT} \right)^2},
\label{eqn:fisher}
\end{equation}
where $\mathbf{s} = \{f_{\rm nl}^y,\alpha\}$ is the set of
parameters to be determined from the data, and the
partial derivatives are evaluated under the null hypothesis
$f_{\rm nl}=0$ and $\alpha = 1$.  The noise with which $f_{\rm nl}$ can be
determined, after marginalizing over $\alpha$, is then
$\left[\left(F^{-1} \right)_{f_{\rm nl}^y f_{\rm nl}^y}
\right]^{1/2}$ and the signal-to-noise $(S/N)$ is $f_{\rm nl}^y$
divided by this quantity.  Numerically, we find $(S/N) \simeq
(f_{\rm nl}^y/1100)( \VEV{y} / 4\times 10^{-9})$.  Thus, the
marginalization over the ISW-tSZ effect only slightly decreases 
the detectability.

Since the noise is again dominated by the tSZ 1-halo term, we can 
perform a similar cleaning to low-redshift sources to that used 
in Section \ref{sec:sz}. If we remove all $z<0.3$ clusters, we 
find that we can detect a smaller value of $f_{\rm nl}$. 
Numerically, using the Fisher matrix as above to marginalise 
over uncertainty in the $yT$ amplitude, we find
$(S/N) \simeq (f_{\rm nl}^y/400)( \VEV{y} / 4\times 10^{-9})$,
closer to the value estimated in Ref.~\cite{Emami:2015xqa}.

\section{Conclusion}
\label{sec:concl}

Here we have calculated the tSZ-ISW cross-correlation,
investigated its use in constraining the redshift distribution
of $y$-parameter fluctuations, and evaluated the detectability
of an early-Universe $yT$ cross-correlation.  We showed that
measurement of the $yT$ cross-correlation can be used to
constrain the redshift distribution of the sources of
$y$-parameter fluctuations, as long as low-redshift tSZ clusters
can be masked, something that may be of utility
given uncertainties in the cluster-physics and
large-scale-structure ingredients (pressure profiles, halo
biases, mass functions) that determine these fluctuations.  We
also showed that estimates, that neglect the $yT$
correlations induced at late times, of the detectability of
early-Universe $yT$ correlations may be optimistic by factors of
a few.

\begin{acknowledgments}
We thank Liang Dai, Yacine Ali-Ha\"{i}moud, and Ely
Kovetz for useful discussions. We also thank the anonymous referee 
for a very quick and conscientious report. SB was supported by NASA through
Einstein Postdoctoral Fellowship Award Number PF5-160133.  This
work was supported by NSF Grant No.\ 0244990, NASA NNX15AB18G,
the John Templeton Foundation, and the Simons Foundation.
\end{acknowledgments}

\end{document}